\documentclass[letter,longauth]{aa}%

\newcommand{\hii}{H\,{\scriptsize II}}

\newcommand{\av}{A$_{\rm V}$}
\usepackage{epsfig,graphicx,color}
\usepackage{ctable}
\usepackage{txfonts}
\begin{document}
\bibliographystyle{aa}
\title{Cluster-formation in the Rosette molecular cloud at the junctions of filaments}
  \author{N. Schneider \inst{1}
  \and T. Csengeri \inst{2}
  \and M. Hennemann \inst{1}
  \and F. Motte \inst{1}
  \and P. Didelon \inst{1}
  \and C. Federrath \inst{3,4}
  \and S. Bontemps \inst{5}
  \and J. Di Francesco \inst{6}
  \and D. Arzoumanian \inst{1}
  \and V. Minier \inst{1}
  \and Ph. Andr\'e \inst{1}
  \and T. Hill \inst{1}
  \and A. Zavagno \inst{7}
  \and Q. Nguyen-Luong \inst{1}
  and 
  \and M. Attard \inst{1}
  \and J.-Ph. Bernard \inst{8} 
  \and D. Elia \inst{9}
  \and C. Fallscheer \inst{6}
  \and M. Griffin \inst{10} 
  \and J. Kirk \inst{10} 
  \and R. Klessen   \inst{4}
  \and V. K\"onyves \inst{1} 
  \and P. Martin \inst{11} 
  \and A. Men'shchikov \inst{1} 
  \and P. Palmeirim \inst{1} 
  \and N. Peretto \inst{1}  
  \and M. Pestalozzi \inst{9} 
  \and D. Russeil \inst{7}
  \and S. Sadavoy \inst{12}
  \and T. Sousbie \inst{13}
  \and L. Testi \inst{14}
  \and P. Tremblin \inst{1}
  \and D. Ward-Thompson \inst{10} 
  \and G. White \inst{15,16}   
     }
 \institute{
  % 1 
  IRFU/SAp CEA/DSM, Laboratoire AIM CNRS - Universit\'e Paris 
  Diderot, 91191 Gif-sur-Yvette, France
  \and 
  % 2
  Max-Planck Institut f\"ur Radioastronomie, Auf dem H\"ugel, 
  Bonn, Germany 
  \and
  % 3  
  Monash Centre for Astrophysics (MoCA), School of Mathematical Sciences, Monash 
  University, Victoria 3800, Australia
  \and
  % 4
  Zentrum f\"ur Astronomie der Universit\"at Heidelberg, 
  Inst. f\"ur Theor. Astrophysik, 
  Albert-Ueberle Str. 2, 69120 Heidelberg, Germany
  \and
  % 5 
  OASU/LAB-UMR5804, CNRS, Universit\'e Bordeaux 1, 33270 Floirac, France    
  \and
  % 6
  National Research Council of Canada, Herzberg Institute of Astrophysics,   
  Victoria BC, Canada 
  \and
  % 7
  Laboratoire d'Astrophysique de Marseille, CNRS/INSU - Universit\'e de Provence, 
  13388 Marseille cedex 13, France
  \and
  % 8 
  Universit\'e de Toulouse, UPS, CESR, 9 av. du colonel Roche, 31028 Toulouse 
  \and 
  % 9 
  IAPS-INAF, Fosso del Cavaliere 100, 00133 Roma, Italy
  \and 
  % 10 
  Cardiff University School of Physics and Astronomy, Cardiff, UK
  \and 
  % 11 
  CITA \& Dep. of Astronomy and Astrophysics, University of Toronto, Toronto, CA
  \and 
  % 12
  Department of Physics and Astronomy, University of Victoria, PO Box 355, 
  STN CSC, Victoria BC Canada 
  \and 
  % 13 
  Institut d'Astrophysique de Paris, UPMC, UMR 7095, CNRS, 98 bldv. Arago, 75014 Paris
  \and 
  % 14
  ESO, Karl Schwarzschild Str. 2, 85748, Garching, Germany
  \and 
  % 15
  Department of Physics \& Astronomy, The Open University, Milton Keynes MK7 6AA, UK 
  \and 
  % 16
  The Rutherford Appleton Laboratory, Chilton, Didcot, OX11 0NL, UK 
  }

%\offprints{}

%\mail{nschneid@cea.fr}

\titlerunning{Cluster formation in Rosette}
\authorrunning{N. Schneider}

\date{\today}

%\date{Received September 15, 1996; accepted March 16, 1997}

\abstract 
% context heading (optional) % 
%
{}
% aims heading (mandatory)
%
{For many years feedback processes generated by OB-stars in molecular
clouds, including expanding ionization fronts, stellar winds, or
UV-radiation, have been proposed to trigger subsequent star formation.
However, hydrodynamic models including radiation and gravity show that
UV-illumination has little or no impact on the global dynamical
evolution of the cloud.  Instead, gravitational collapse of filaments
and/or merging of filamentary structures can lead to building up dense
high-mass star-forming clumps. However, the overall density structure
of the cloud has a large influence on this process, and requires a
better understanding.}
%
% methods heading (mandatory)
{The Rosette molecular cloud, irradiated by the NGC~2244 cluster, is a
template region for triggered star-formation, and we investigated its
spatial and density structure by applying a curvelet analysis, a
filament-tracing algorithm (DisPerSE), and probability density
functions (PDFs) on Herschel\thanks{Herschel is an ESA space
observatory with science instruments provided by European-led
Principal Investigator consortia and with important participation from
NASA.} column density maps, obtained within the HOBYS key program.}
%
% results heading (mandatory)
{The analysis reveals not only the filamentary structure of the cloud
but also that all known infrared clusters except one lie at
junctions of filaments, as predicted by turbulence simulations.  The
PDFs of sub-regions in the cloud show systematic differences. The two
UV-exposed regions have a double-peaked PDF we interprete as caused by 
shock compression, while the PDFs of the center and other cloud parts
are more complex, partly with a power-law tail. A deviation of the
log-normal PDF form occurs at \av $\approx$9$^m$ for the center, and
around 4$^m$ for the other regions.  Only the part of the cloud
farthest from the Rosette nebula shows a log-normal PDF.  }
%
% conclusions (optional)
{The deviations of the PDF from the log-normal shape typically
associated with low- and high-mass star-forming regions at \av
$\approx$3--4$^m$ and 8--10$^m$, respectively, are found here within
the very same cloud. This shows that there is no fundamental
difference in the density structure of low- and high-mass star-forming
regions. We conclude that star-formation in Rosette -- and probably in
high-mass star-forming clouds in general -- is not globally triggered
by the impact of UV-radiation. Moreover, star formation takes place in
filaments that arose from the primordial turbulent structure built up
during the formation of the cloud. Clusters form at filament mergers,
but star formation can be locally induced in the direct interaction
zone between an expanding \hii-region and the molecular cloud.}

\keywords{interstellar medium: clouds
          -- individual objects: Rosette 
           }

   \maketitle

%________________________________________________________________

\begin{figure*}[ht]
\begin{center} 
\hspace{-0.5cm}
\includegraphics[angle=0,width=8.3cm]{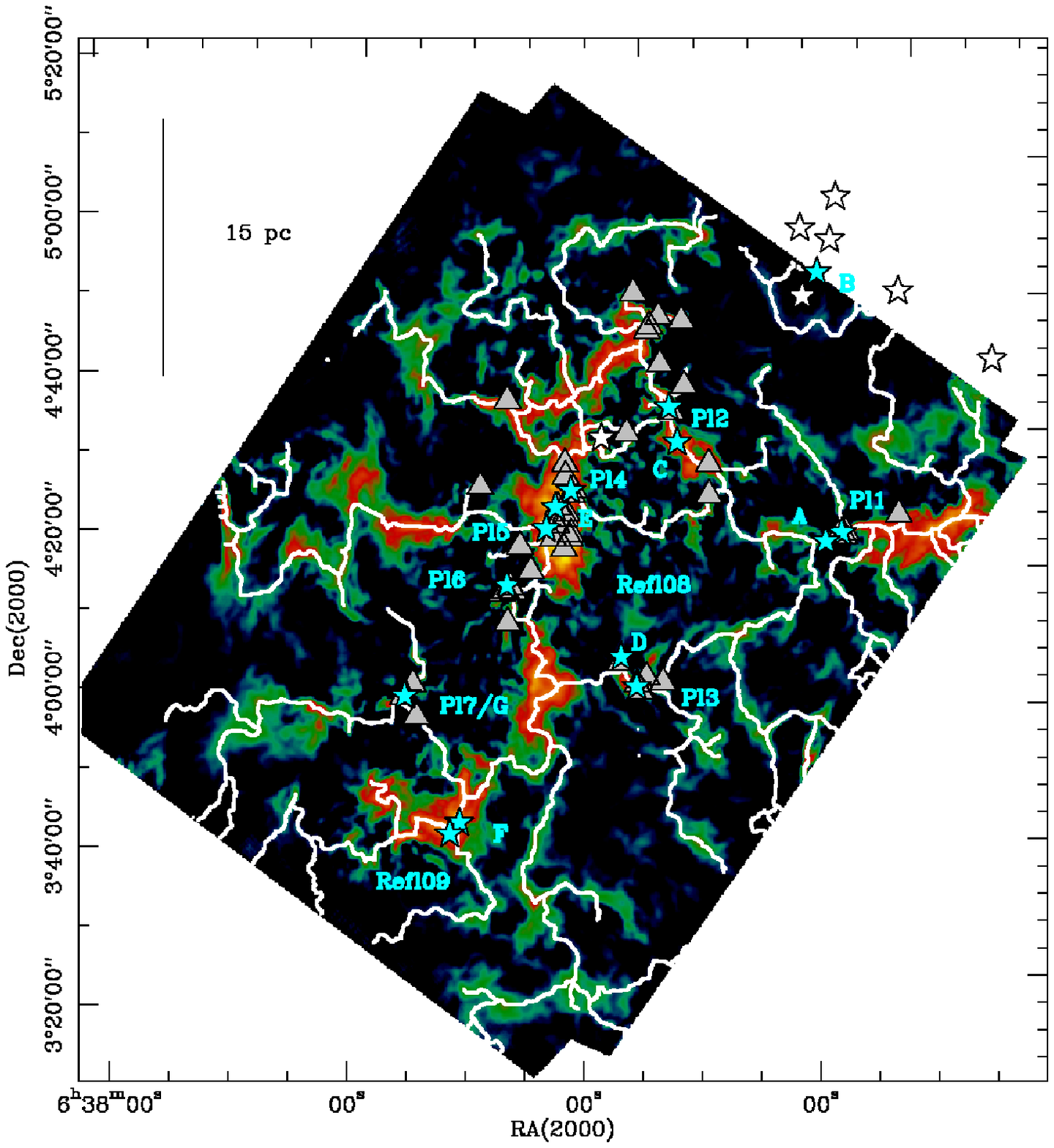}
\hspace{0.5cm}
\vspace{0.5cm}
\includegraphics [width=9.0cm, angle={0}]{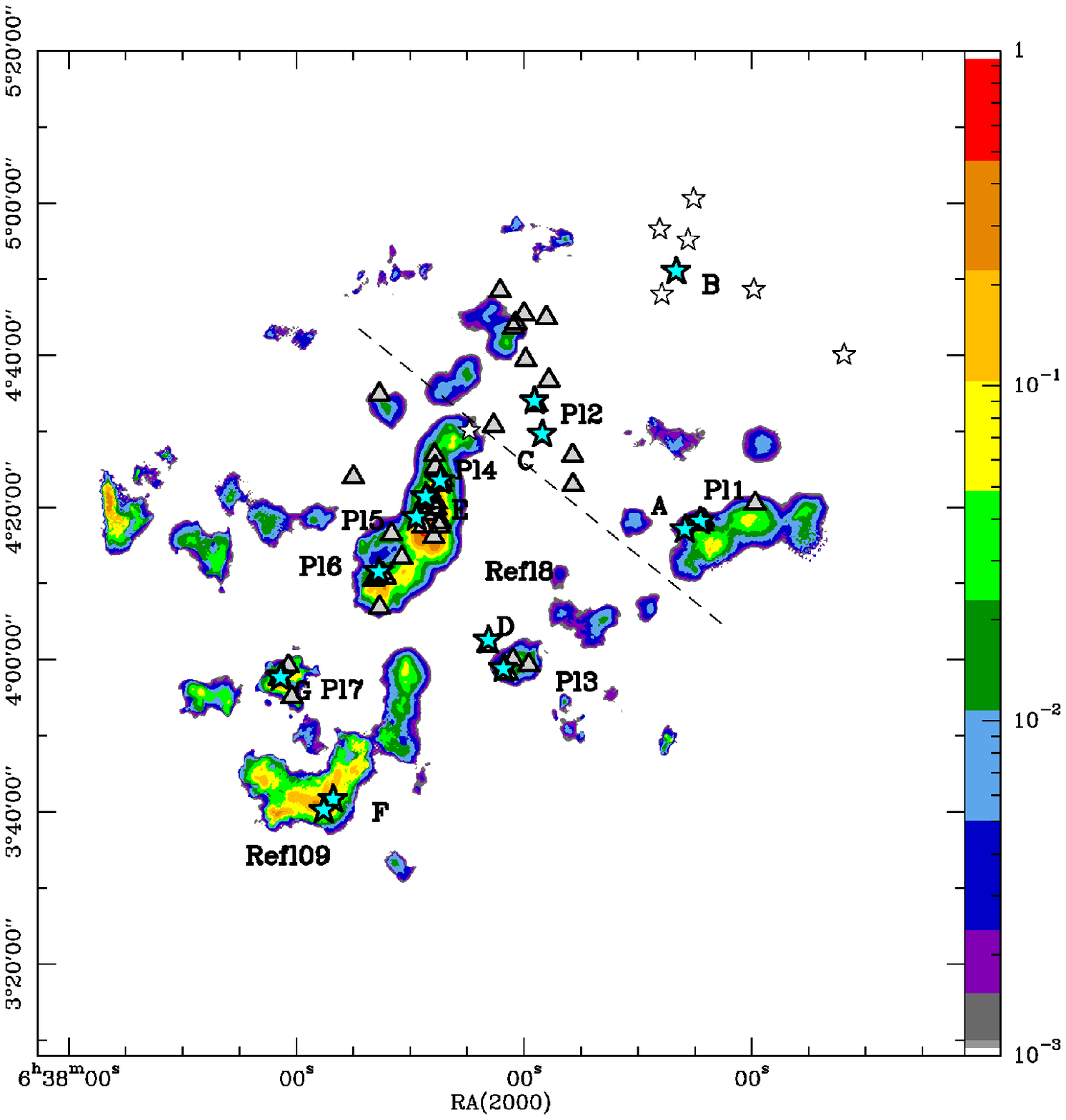}
\caption [] {{\bf Left:} Curvelet map of the column density, expressed
in logarithmic magnitudes of visual extinction (\av = 0.05 to 7).  The
absolute values, however, depend on the decomposition threshold
($\approx$20\% of the intensity are in the curvelets) and are not
relevant for this study. The filamentary structure becomes apparent
and is additionally traced by DisPerSE (white lines). See Appendix B for
more details and a close-up of the central region.  Known
infrared-clusters are indicated as blue stars ('Pl' for Phelps \& Lada
\cite{phelps1997}, A--F for Poulton et al. \cite{poulton2008}, and
Refl for Rom\'an-Z\'u\~niga et al. 2008). White stars in the upper right
corner indicate the O-stars from NGC~2244. Gray triangles are the
massive dense cores (MDCs) identified in Motte et
al. (\cite{motte2010}).  {\bf Right:} A map showing the 'confidence' 
(coded from 0 to 1 in color scale) of finding cluster formation sites.  This map
combines the information of the column density map and the locations
of filament junctions (see text for details). Known clusters and MDCs
are labeled as in the left panel. The dashed line indicates the
approximate border beyond which (toward the cluster) SF due to the direct 
compression scenario may be at work.}
\label{skeleton}
\end{center} 
\end{figure*}

\section{Introduction} \label{intro}

%Feedback processes of massive stars 
The role of feedback from massive stars has been long discussed as a
dynamical trigger for subsequent star formation (SF).  One has to
distinguish, however, between the driving scales on which the
processes work -- and their relative importance.  {\sl Large-scale}
turbulence, even on a Galactic scale (Mac Low \& Klessen
\cite{maclow2004}), can be excited by expanding supernova shells, 
colliding flows forming clouds, or  
globally by any form of accretion (Klessen \& Hennebelle
\cite{klessen2010}).  {\sl Intermediate- to small-scale} processes
include radiation, stellar winds, and outflows. Observationally,
expanding ionization fronts were proposed to account for fragmented
shells around bubble-like \hii-regions (e.g. Zavagno et
al. \cite{zavagno2010}).  Shock compression of a simple layer is the
initial idea of the {\sl collect and collapse scenario} (Elmegreen \&
Lada \cite{elmlada1977}). Recent models (Walch et
al. \cite{walch2011}) consider now an initially inhomogeneous
cloud. In particular Dale \& Bonnell (\cite{dale2011},
\cite{dale2012}) showed that photoionizing
radiation from an OB-cluster on a spatially highly inhomogeneous
molecular cloud has little or no impact on the global evolution of the
cloud due to the presence of strong accretion flows. Generally, in a
dynamic view of molecular cloud- and SF-interactions (e.g.,
Ballesteros-Paredes et al.  \cite{ball2007}), stars form in
gravitationally collapsing dense filaments that formed in
turbulence-generated, interacting shocks. Dale \& Bonnell 
(\cite{dale2011}) proposed that the most massive clusters should be
found at the {\sl junctions of filaments}.  This confirms the finding
of Schneider et al. (2010a) where a mass flow through subfilaments,
probably governed by magnetic fields, was observed for the DR21
filament. The prediction that (massive) clusters form preferentially
in the deep potential wells where filaments merge can be verified
observationally and it is within the scope of this paper to do so. We
chose the Rosette Molecular cloud (RMC) as a 
proposed template for 'triggered' SF. A discussion of the advantages and 
disadvantages of this premise can be found in Schneider et al. (2010b).  To
characterize the cloud structure, a curvelet decomposition of column
density maps obtained with {\it Herschel} data is performed, and a
filament-tracing algorithm is applied to the map (see Arzoumanian et
al. 2011 and Appendix B for details). \\
%
%Rosette 
The Rosette nebula at a distance of 1.6 kpc is illuminated by the
central OB cluster \object{NGC~2244} located inside a cavity.  The expanding
\hii-region is interacting with a GMC (mass of a few $10^5~{\rm
M}_\odot$, Williams et al. \cite{williams1995}). Photon dominated
regions (PDRs) are detected not only along the interface but also deep
inside the molecular cloud (Schneider et al. \cite{schneider1998}),
reflecting the deep penetration of UV-radiation owing to its
inhomogenous structure.
%
%PDFs
We investigated the density structure of the RMC by deriving
probability density functions (PDFs) of the column
density\footnote{expressed in visual extinction \av\, with
N(H$_2$)/\av=0.94$\times$10$^{21}$ cm$^{-2}$ mag$^{-1}$ (Bohlin et
al. \cite{bohlin1978}).} of the whole cloud and subfields in the
Rosette. To do this, we used observations from {\it Herschel} within the
HOBYS ({\it Herschel} imaging survey of OB Young Stellar objects,
Motte, Zavagno, Bontemps et al. 2010) guaranteed time key
program. Column density maps were determined from a modified black
body fit to the wavelengths of PACS and SPIRE (see Appendix A) and
have an angular resolution of $\approx$37$''$.  Generally, a PDF
characterizes the fraction of gas that has a column density $N$ in the
range [$N$, $N$+$\Delta N$] (e.g., Federrath et al.
\cite{fed2008}).

%__________________________________________________________________

\section{Results and analysis} \label{results}

\subsection{The filamentary structure of Rosette} \label{skl} 

Figure~\ref{skeleton} (left) shows a curvelet image from a
multi-resolution morphological decomposition (Starck et
al. \cite{starck2004}) of the column density map of the Rosette (see
Fig. 2) obtained from {\it Herschel} data.  Overlaid on the image are
the filaments traced by the DisPerSE algorithm (Sousbie et
al. \cite{sousbie2011}). This method reveals the highly filamentary
structure of the RMC, which is not fully apparent from visual inspection of the
column density map (or molecular line maps). Existing infrared (IR)
clusters and the most massive dense cores (MDCs) identified in the
same data set (Motte et al. 2010; Hennemann et
al. \cite{hennemann2010}) are overlaid on the image. The latter are
potential sites of future massive star formation. It is apparent that
all sources lie in the proximity of junctions of the most prominent
filaments in high column density regions. Some of the MDCs and the
cluster Pl2, however, are also located along filaments (see, e.g.,
close-up of the center region shown in Fig. 5).  To better
qualitatively characterize the filament-junction/cluster correlation,
we created a 'confidence map' that combines the two requirements for
cluster formation: high column density and filament junction. First,
we produced a mask containing only the junction points and then
defined a Gaussian distribution function with a FWHM\footnote{The
radii of the clusters vary between 1.3 and 3.6 pc (Rom\'an-Z\'uniga
et al. 2008), as a conservative value we assumed 1.5 pc.  However, the
resulting map does not change much in a range of $\sim$1 to $\sim$5
pc.} of 3$'$=1.5 pc around each point. This procedure yields a
'junction-area' that is more realistic than only a single point. We
then multiplied this map with the column density map, and normalized
the resulting map to obtain values between 0 and 1.  Inspection of
Fig.~\ref{skeleton} (right) shows that the IR-clusters and many MDCs
are indeed found in regions of high column density where filaments
merge (all within the maximum cluster size of 7 pc). The only
exception is the \hii-region/molecular cloud interface (northwest of
the dashed line in Fig.~\ref{skeleton}). Neither cluster Pl2 nor the
majority of MDCs lie in regions of high confidence. Here, shock
compression of the expanding ionization front with subsequent
fragmentation into dense cores may be at work, as suggested by
the PDFs of this region (see next Sec.~\ref{pdf_rosette}).

\begin{figure}[ht]
\begin{center} 
\hspace{-0.5cm}\includegraphics[width=7cm, angle={0}]{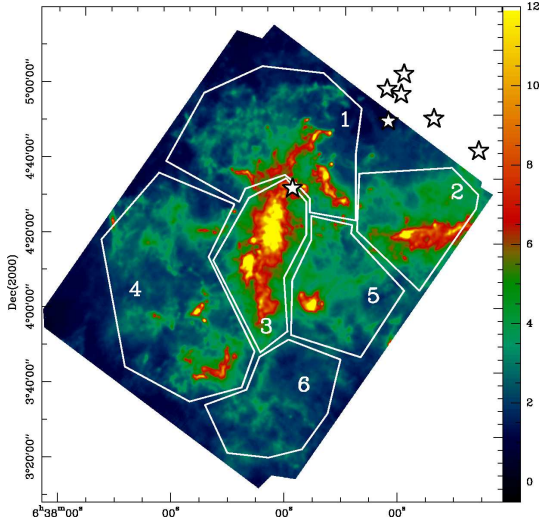}
\end{center} 
\caption [] {Column density map of the Rosette molecular
cloud, expressed in visual extinction (see color bar). The six regions that were
selected to determine individual PDFs are indicated by polygons. }
\label{rosette-herschel-pdf}
\end{figure}

\subsection{PDFs from the Herschel column density map of Rosette} \label{pdf_rosette} 

\begin{figure}[ht]
\hspace{-0.0cm}\includegraphics[angle=90,width=4.0cm]{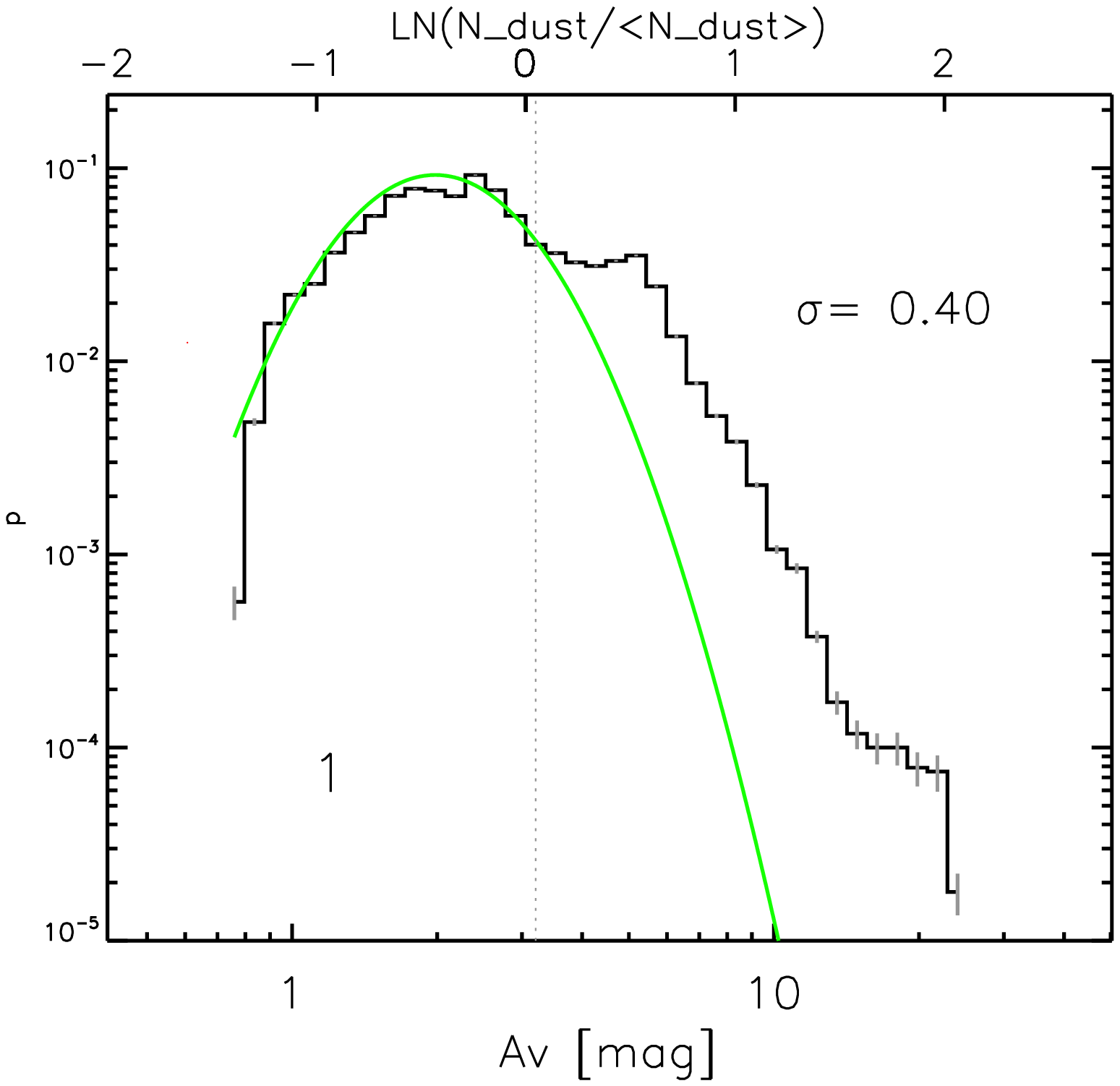}
\hspace{-1.0cm}\includegraphics[angle=90,width=4.0cm]{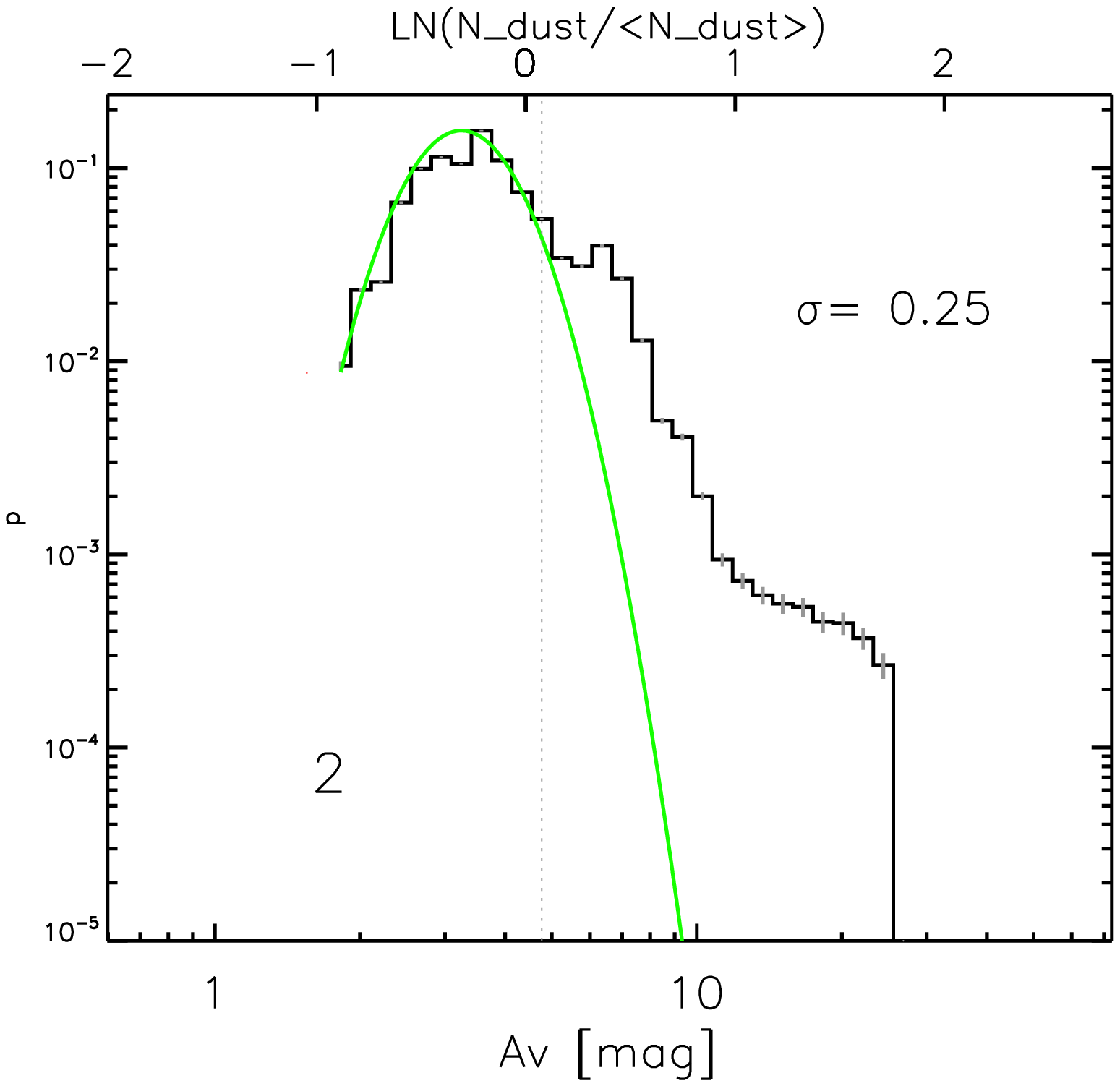}
\vspace{0.3cm}
\hspace{-0.0cm}\includegraphics[angle=90,width=4.0cm]{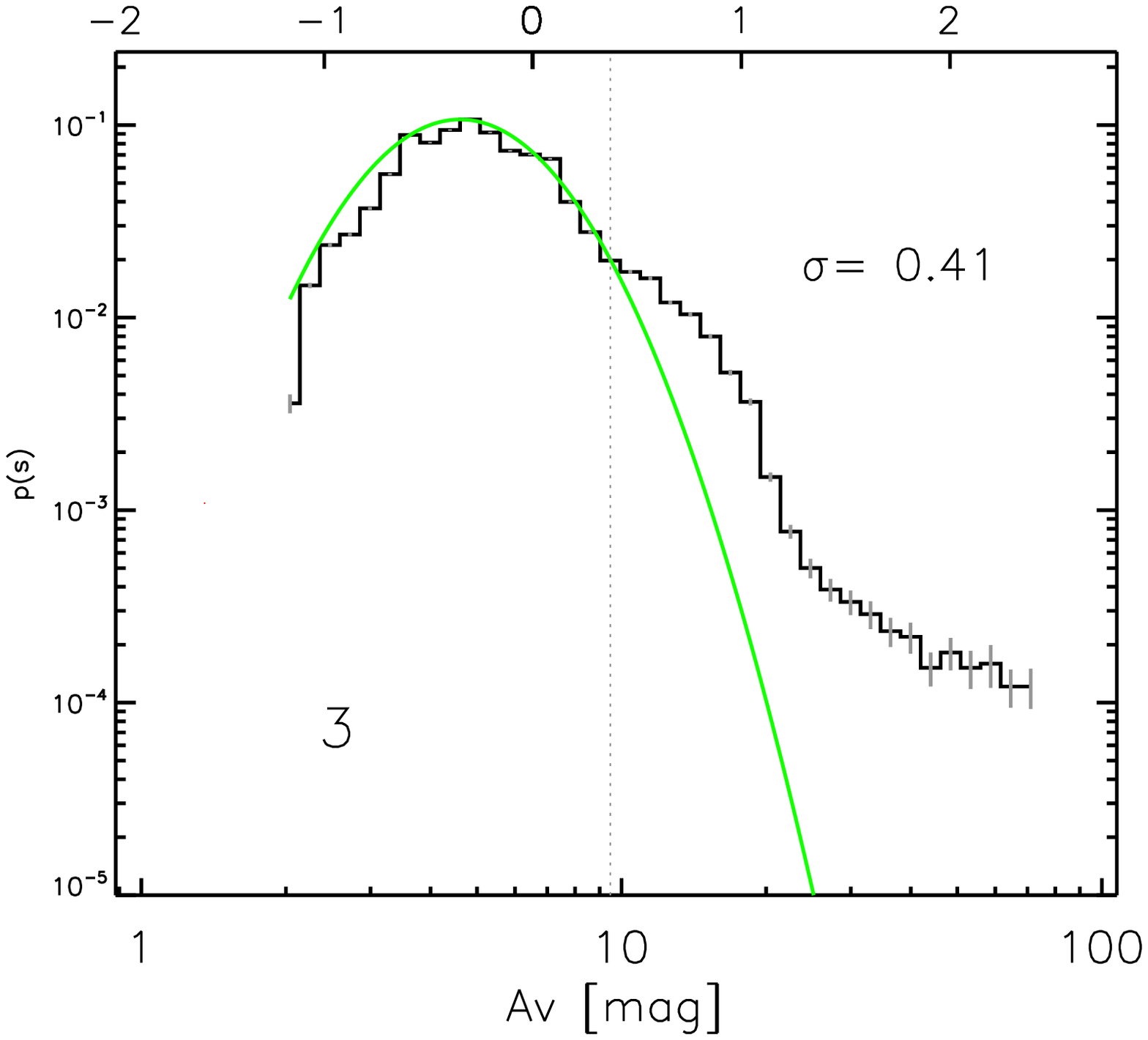}
\hspace{-1.0cm}\includegraphics[angle=90,width=4.0cm]{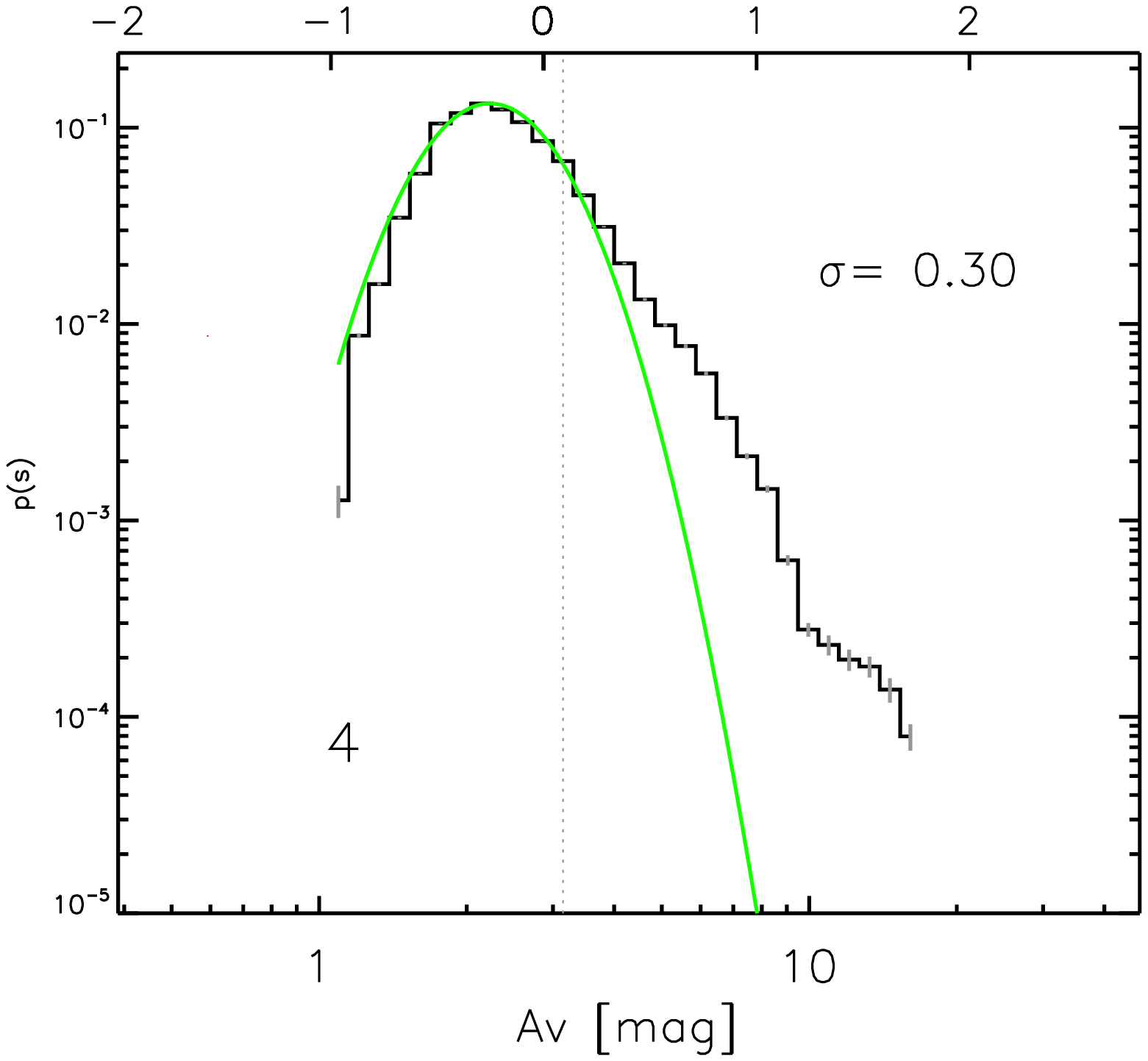}
\vspace{0.3cm}
\hspace{-0.0cm}\includegraphics[angle=90,width=4.0cm]{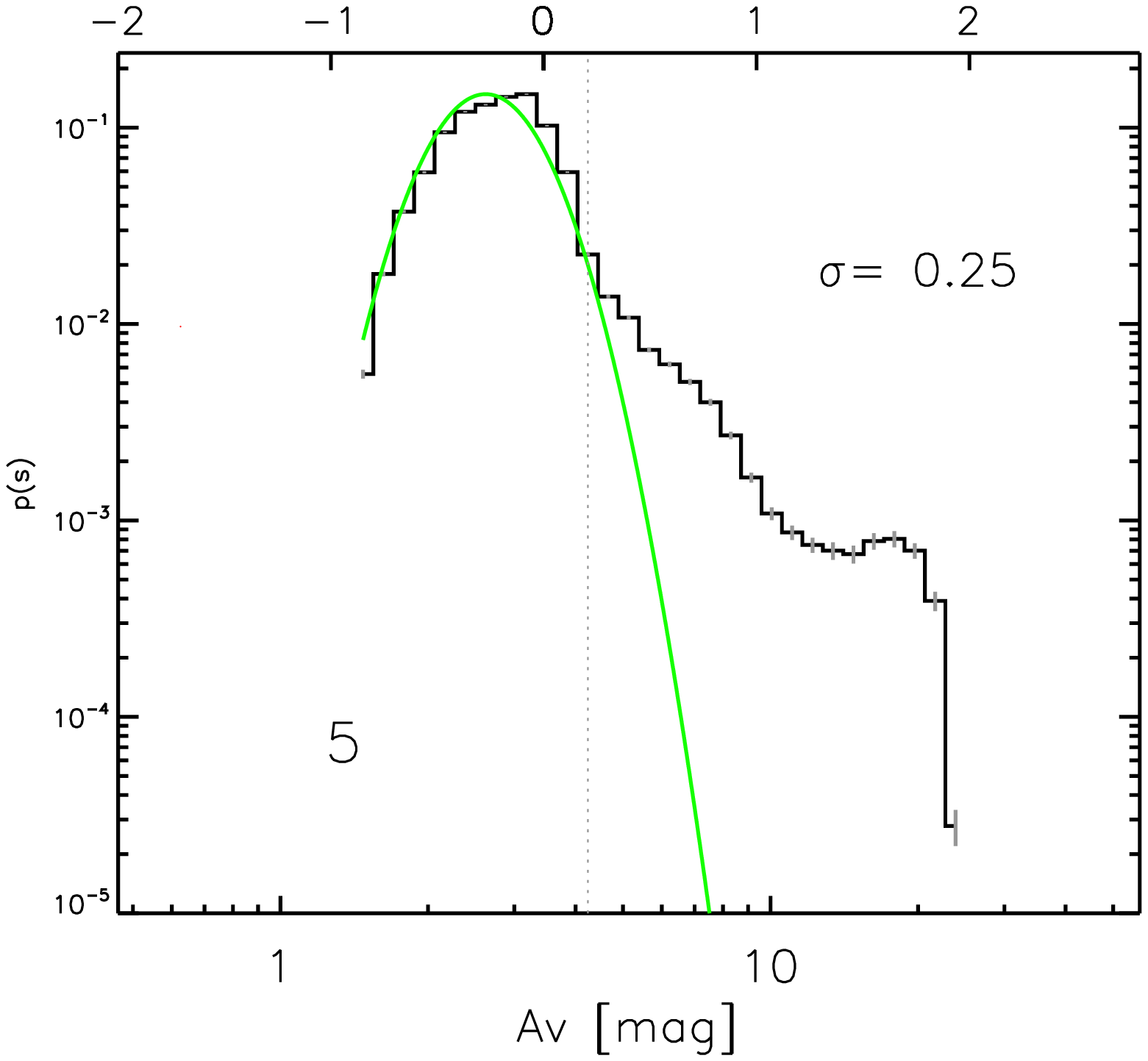}
\hspace{-0.2cm}\includegraphics[angle=90,width=4.0cm]{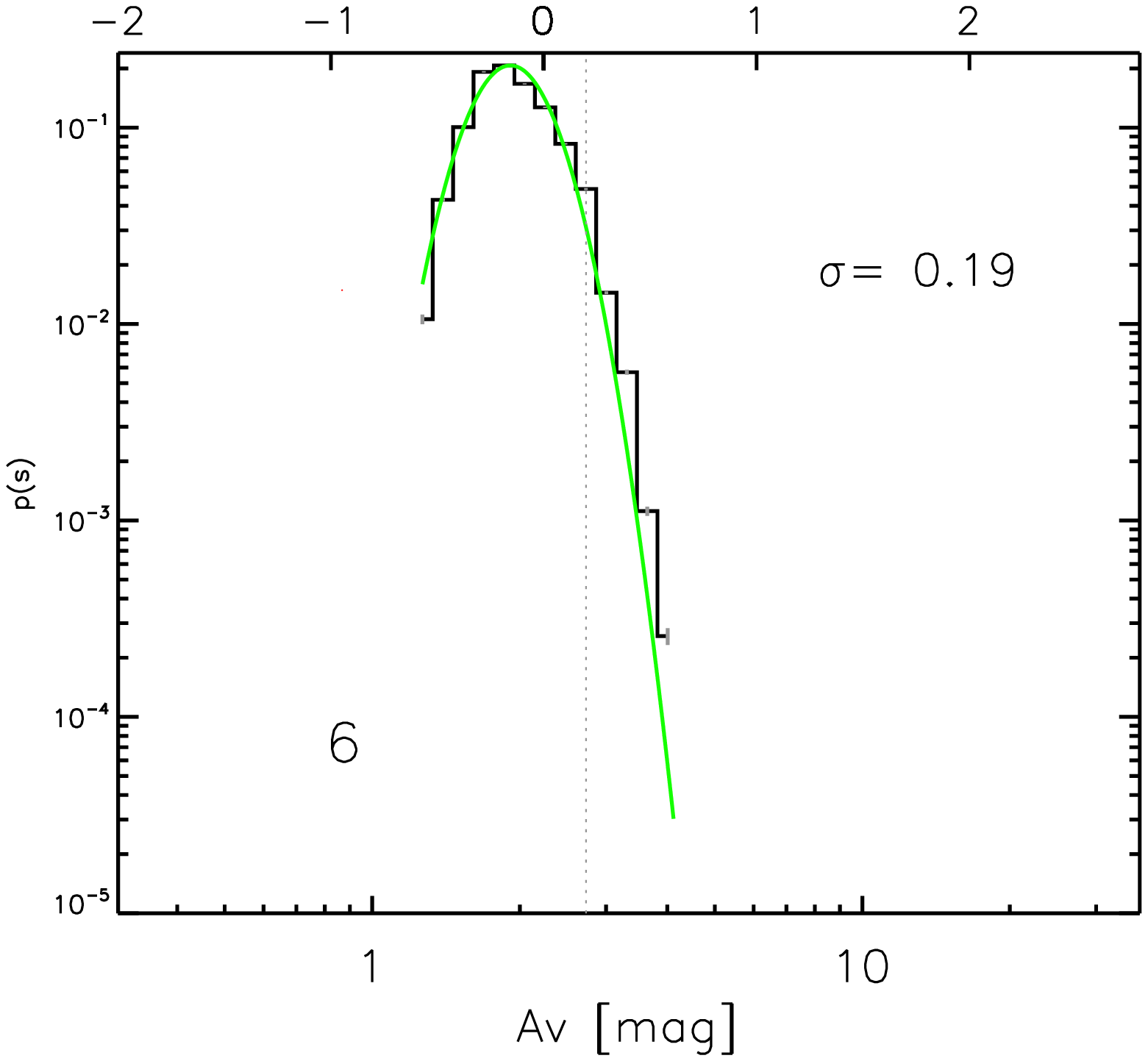}
\caption [] {PDFs for the six subregions in the Rosette, labeled from 1 to 6. 
The green line indicates the fitted PDF. The \av\,-value of the
deviation of the PDF from the log-normal form is shown as a dotted
vertical line and the width of the PDF ($\sigma_{\eta}$) is given in
each panel.}
\label{pdf-subregions}
\end{figure}

Figure~\ref{rosette-herschel-pdf} shows the column density map
including our division into several subregions, reflecting different
morphological and physical properties. {\bf (1)} and {\bf (2)}:
interface of \hii-region and molecular cloud with many UV-exposed
pillars, {\bf (3)}: dense, main SF region of the RMC, including
several IR-clusters (Pl4/5/6), {\bf (4)} and {\bf (5)}: more diffuse
and colder gas with dense, SF clumps (Pl7 in region 4 and Pl3 in
region 5), {\bf (6)}: quiescent and cold remote part of the cloud
without SF activity.

The PDF obtained for the whole molecular cloud is displayed in 
Appendix C (Fig. 6). A log-normal PDF with a width of
$\sigma_{\eta}$=0.63 is fitted up to \av\,=3$^m$ (noise level
$\approx$1$^m$).  The excess at high column densities
(\av\,=3--20$^m$) follows a power law. The slope corresponds to a
volumetric, radial density profile of $n
\propto r ^{-0.65}$ (Federrath et al. \cite{fed2011}). This does not 
assume that the cloud is a single sphere, moreover it consists of 
many clumps/cores, each with a radial profile leading to a
common exponent $\alpha$. The exponent is smaller than what is 
typically found for dense cores (--1.5 to --2), suggesting that on
large (cloud)-scales, turbulence is the dominating process compared to
gravity.

Figure~\ref{pdf-subregions} displays the PDFs for the six subregions,
all showing a log-normal distribution at low \av\, and a more complex
curvature or several peaks at higher \av.  Particularly interesting
are the PDFs of the UV-exposed regions 1 and 2 (the OB-cluster is only
10--15 pc away in projected distance) because there are two 'peaks'
(at \av $\approx$3$^m$ and 6$^m$). A similar double-peak PDF was
observed (Schneider et al., priv. comm.) using a column density map of
RCW120 (Zavagno et al. 2010), a very good example of a simple,
'bubble'-like geometry. We interpret the second peak as arising from a
higher column density component due to gas compression by the
expanding ionization front. This interpretation is supported by recent
numerical models including turbulence and radiation (Tremblin et al.,
in prep.) that exhibit double-peak PDFs depending on the turbulent
state of the cloud and the initial curvature of the cloud surface. We
therefore do not expect to see this feature in all UV-illuminated
environments.  Region 3 is the central SF region with the highest
column density (up to 70$^m$). The PDF here is broad
($\sigma_{\eta}$=0.41), shows two 'breaks' (around 9$^m$ and 20$^m$),
and has a peak value at
\av\,$\approx$5$^m$ (all other PDFs peak at $\approx$2$^m$). The PDFs
of regions 4 and 5 cover a lower density regime and deviate from the
log-normal form around \av\,$\approx$4$^m$. The most quiescent region
6 is the only one displaying a well-defined log-normal PDF with the
smallest width ($\sigma_{\eta}$=0.19).

\section{Discussion} \label{discuss} 
\subsection {The density structure of the RMC} 
\noindent 
In Csengeri et al. (in prep.), we found that low-mass SF regions have
a break in the log-normal shaped PDF at \av\,$\approx$4$^m$ (and
possibly around 8$^m$), and high-mass SF regions around
\av\,$\approx$8$^m$ and higher. We now show that these two breaks in \av\, are
found in the PDFs of subregions within one {\sl single cloud}, the
Rosette GMC, suggesting that there is {\sl no fundamental difference}
in the density structure of low-mass and high-mass SF regions up to an
\av\, of typically 7--10$^m$. Above this value, the tail(s) in the
PDFs are associated with SF activity. Namely, if gravity starts to
play a role during molecular cloud formation, an increasing fraction
of gas will exceed a certain threshold of column density/extinction
and form stars. Other studies (e.g., Lada et al. \cite{lada2010};
Heiderman et al. \cite{heider2010}; Andr\'e et al., 2010) have arrived
at the same value of around 8$^m$. In addition, only numerical models
that include gravity are able to reproduce this threshold and the
corresponding break in the PDF. For example, Ballesteros-Paredes et
al. (2011) showed that PDFs vary during cloud evolution.  Purely
log-normal shapes were found in an initially turbulent,
non-gravitating cloud, while one or more log-normal PDFs at low
column-densities and a power-law tail for higher values were found for
later stages of cloud evolution.  In addition, the slope of the
power-law tails varies with time. All PDFs in the Rosette have the
shape of the {\sl later evolutionary states}, indicating that gravity
remains an important process in cloud evolution, including
star-formation. The impact of UV radiation, however, needs to be
investigated using hydrodynamic simulations including gravity and
radiation.

\subsection{Where do stars form  ?} 
\noindent 
Our Rosette observations indicate that clusters preferentially form at
{\sl filament junctions} but cores (and subsequently stars) can also
form {\sl along filaments} (see also other studies of high-mass SF
regions, e.g., the Nessie-nebula (Jackson et al. \cite{jackson2010}),
W48 (Nguyen Luong et al. \cite{quang2011}) or Cygnus (Motte et
al. \cite{motte2007})).  In {\it Herschel} observations of low-mass SF
regions (Andr\'e et al. \cite{andre2010}; Arzoumanian et
al. \cite{doris2011}) dense cores were found to be located along
super-critical filaments\footnote{gravitationally unstable with a mass
per unit length larger than 2\, c$_s^2$/G with the isothermal sound
speed c$_s$} without higher
concentrations at the junctions.  This points toward a scenario in
which it is the large potential well of merging filaments
(subcritical or supercritical) that enables the formation of massive
stars within a cluster. Conform with numerical simulations
(e.g. Klessen \& Hennebelle \cite{klessen2010}; V\'azquez-Semadeni et
al. \cite{vaz2009}) it is therefore ultimately the total mass and the high
accretion rate that is the decisive factor for the formation of a
cluster. Observationally, apart from the present study, there are more
examples of how massive structures can be built up from parsec-scale
filaments (DR21, Schneider et al. \cite{schneider2010a} and Hennemann
et al., in prep.; Vela C, Hill et al. 2011). The supply of material
may continue on much smaller (subparsec) scales to build-up a cluster
with massive stars, as proposed by, e.g., Csengeri et
al. (\cite{csengeri2011}). Alternatively, massive stars may form in a
turbulence-regulated, quasi-static scenario (McKee \& Tan
\cite{tan2003}). However, the discussion of these small, subparsec
scale processes is beyond the scope of this paper.

\subsection{Does triggered star-formation exist ? } 

Dale \& Bonnell (2011, 2012) simulated the gravitational collapse of a
turbulent (giant) molecular cloud exposed to photoionizing
radiation. Their main conclusion was that although the gas and dust is
heated, photoionization has no strong {\sl global}, i.e. tens of
parsec scale, effect on the dense gas because the massive clumps out
of which OB clusters form are continuously accreting and most of the
photoionizing flux is absorbed by these accretion flows.
Observationally, no clear picture emerges. In the RMC, a very low
UV-field in the remote part of the cloud was determined ($\approx$2-8
Habing at a distance of 30 pc, Schneider et al. 1998), supporting the
idea of low UV-impact.  On the other hand, a temperature gradient, and
a tentative age-gradient of sources, was derived (Schneider et
al. 2010b)\footnote{Temperature gradients on smaller scales ($<10$pc)
were also seen in M16 (Hill et al. \cite{hill2012}, submitted) and Vela C (Minier et
al., in prep.)}. In addition, {\sl local triggering} of SF, as seen
in the RMC (and RCW36, Minier et al., in prep.) is possible
(Sec. 2.1). The observed double-peak PDFs (Sec. 2.2) support this
scenario. We therefore conclude that though the morphological and PDF
analysis of Rosette shows no evidence that SF is {\sl globally
triggered} (across the whole extent of the cloud), but is moreover
governed by gravity, the impact of UV-radition is not settled and
needs more investigation.

\begin{acknowledgements}
We thank V. Ossenkopf, E. V\'azquez-Semadeni, and J. Dale for fruitful
discussions, and the referee, J. Williams, for useful comments. 
SPIRE has been developed by a consortium of institutes
led by Cardiff University (UK) and including Univ. Lethbridge
(Canada); NAOC (China); CEA, LAM (France); IFSI, Univ. Padua (Italy);
IAC (Spain); Stockholm Observatory (Sweden); Imperial College London,
RAL, UCL-MSSL, UKATC, Univ. Sussex (UK); and Caltech, JPL, NHSC,
Univ. Colorado (USA). This development has been supported by national
funding agencies: CSA (Canada); NAOC (China); CEA, CNES, CNRS
(France); ASI (Italy); MCINN (Spain); SNSB (Sweden); STFC (UK); and
NASA (USA).  PACS has been developed by a consortium of institutes led
by MPE (Germany) and including UVIE (Austria); KU Leuven, CSL, IMEC
(Belgium); CEA, LAM (France); MPIA (Germany); INAF-IFSI/OAA/OAP/OAT,
LENS, SISSA (Italy); IAC (Spain). This development has been supported
by the funding agencies BMVIT (Austria), ESA-PRODEX (Belgium),
CEA/CNES (France), DLR (Germany), ASI/INAF (Italy), and CICYT/MCYT
(Spain).  Part of this work was supported by the ANR (\emph{Agence
Nationale pour la Recherche}) project ``PROBeS'', number
ANR-08-BLAN-0241.  R.S.K. acknowledges support from the the German
{\rm Bundesministerium f\"ur Bildung und Forschung} via the ASTRONET
project STAR FORMAT (grant 05A09VHA) and from the {\em DFG} via grant
SFB 881.  T. Cs. acknowledges financial support for the ERC Advanced
Grant GLOSTAR under contract no. 247078.
\end{acknowledgements}

\Online

\begin{appendix} 

\section{Observations and column density map} \label{obs}

% A.1
\onlfig{4}{
\begin{figure*}[]    
\begin{center} 
\includegraphics[angle=0,width=13cm]{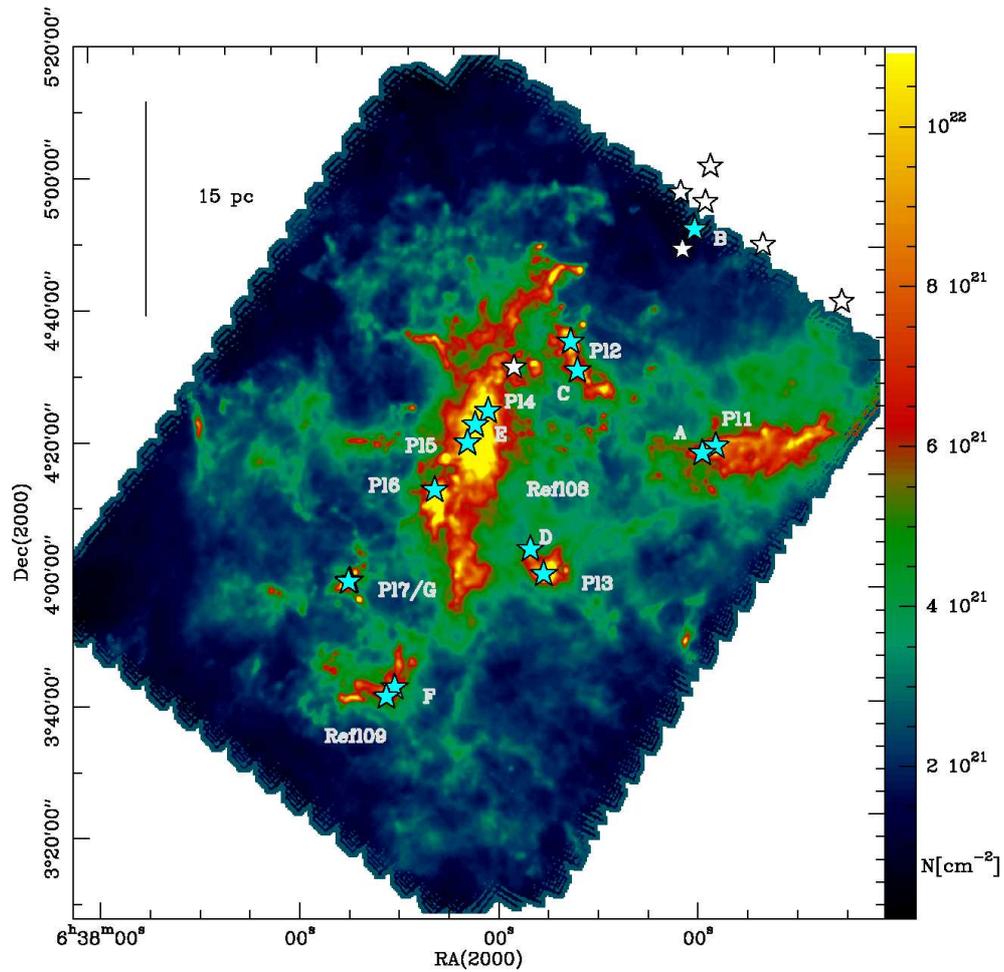}
\end{center} 
\caption [] {Column density map of the Rosette Molecular Cloud, obtained 
from the 160, 250, 350, and 500 $\mu$m maps from {\it Herschel}. 
Known infrared-clusters are indicated in the plot as blue stars ('Pl'
refers to Phelps \& Lada (\cite{phelps1997}), A, B, etc. the clusters 
from Poulton et al. (\cite{poulton2008}), and labeling ``Refl'' those from Rom\'an-Z\'u\~niga et
al. (2008)). White stars in the upper right corner indicate the
O-stars from NGC2244.}
\label{coldens}
\end{figure*}
}

The Rosette was observed during the Science Demonstration Phase (SDP)
within the HOBYS key program. This program is dedicated to the
earliest stages of high-mass star formation and images all 
molecular complexes that form high-mass stars at less than 3 kpc with
SPIRE (Griffin et al. \cite{griffin2010}) and PACS (Poglitch et
al. \cite{poglitsch2010}) using the {\it Herschel} satellite (Pilbratt
et al. \cite{pilbratt2010}).  The SPIRE and PACS data\footnote{Data
are public available in the {\sl Herschel} Science Archive.} from 70
$\mu$m to 500 $\mu$m were obtained on October 20, 2009 in parallel
mode with a scanning speed of 20$''$/sec. Two orthogonal coverages of
size 1$^\circ$45$'\times$1$^\circ$25$'$ were performed, mapping the
(largest) southeast part of the Rosette Molecular Cloud. The SPIRE
data were reduced with HIPE version 7.1956, using a modified version
of the pipeline scripts, i.e., observations that were taken during the
turnaround at the map borders were included, the most recent
calibration tree was used, and the destriper-module with a polynomial
baseline of 0th order was applied. The two orthogonally scanned maps
were then combined using the `naive-mapper' (i.e., a simple averaging
algorithm). The destriper module significantly improved the resulting
maps, compared to the first data reduction just after the SDP
(Schneider et al. 2010b).  The angular resolutions at 160, 250, 350,
and 500 $\mu$m, are $\sim$12$''$, $\sim$18$''$, $\sim$25$''$, and
$\sim$37$''$, respectively.

The column density was determined from a pixel-to-pixel modified black
body fit to four wavelengths of PACS and SPIRE (160, 250, 350, 500
$\mu$m, all maps were smoothed to the beamsize of the 500 $\mu$m map,
i.e., $\sim$37$''$). For the region covered by PACS and SPIRE
simultaneously, we fixed the specific dust opacity per unit mass
(dust+gas) approximated by the power law $\kappa_\nu \, = \,0.1 \,
(\nu/1000 GHz)^\beta$ cm$^{2}$/g and $\beta$=2, and left the dust
temperature and column density as free parameters (see Hill et
al. 2011; Arzoumanian et al. 2011 for details). For the region only
covered with SPIRE\footnote{The instruments are offset by 11$'$ in the
focal plan.} we used 17.3 K, the median value of the SED-derived
temperature across the main map where PACS and SPIRE overlap giving us
four-band coverage at 160, 250, 350, and 500 $\mu$m.  We checked the
fitted SEDs and found no major discrepancy in the fits, though the
method described above assumes a single temperature and optically thin
emission, which is not always a good approximation if there is a
temperature variation along the line-of-sight, noise, and if the dust
becomes optically thick at shorter wavelengths. Shetty et
al. (\cite{shetty2009}), for example, showed that this can even
produce an anti-correlation of $\beta$ and T which was claimed to be
observed by other authors. However, it is beyound the scope of this
paper to go more into detail. While the column density map shown in
Schneider et al. (2010b) was calibrated using extinction maps, we now
recovered the {\it Herschel} zero-flux levels of the Rosette field
with Planck data (Bernard et al., priv. communication). The final
column density map is shown in Fig. 4.

\onlfig{5}{
\begin{figure*}[]
\begin{center} 
\includegraphics [width=7cm, angle={-90}]{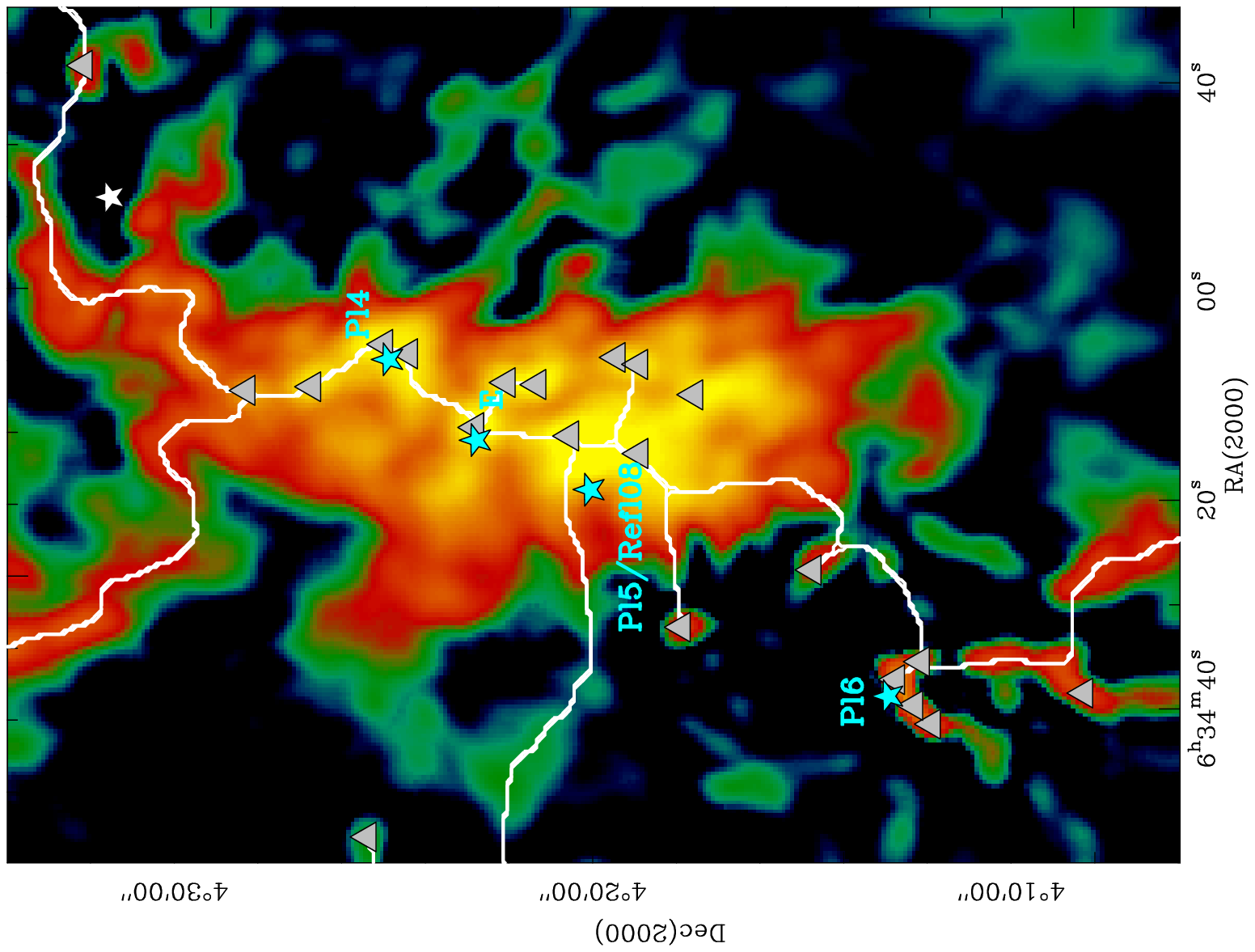}
\includegraphics [width=7cm, angle={-90}]{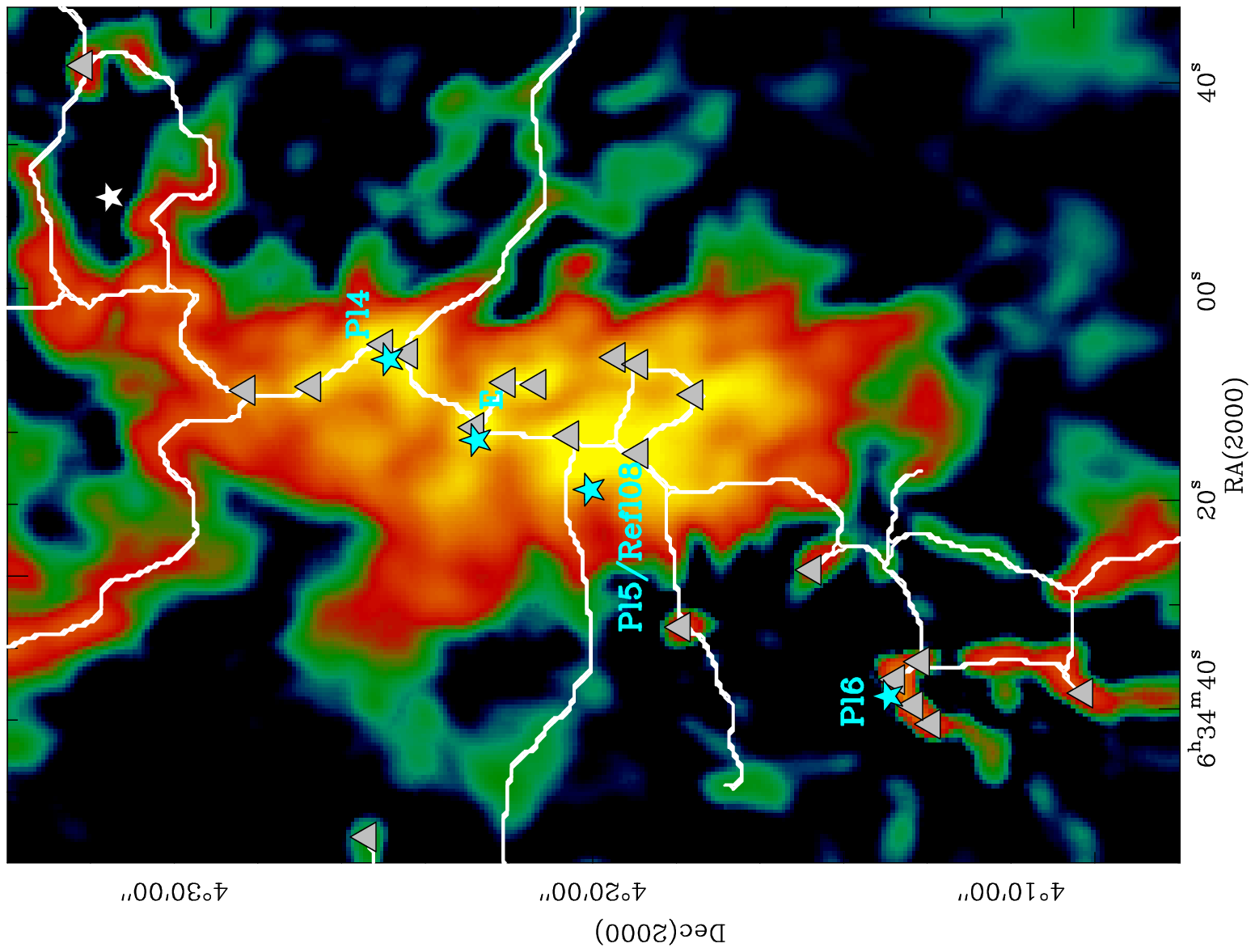}
\includegraphics [width=7cm, angle={-90}]{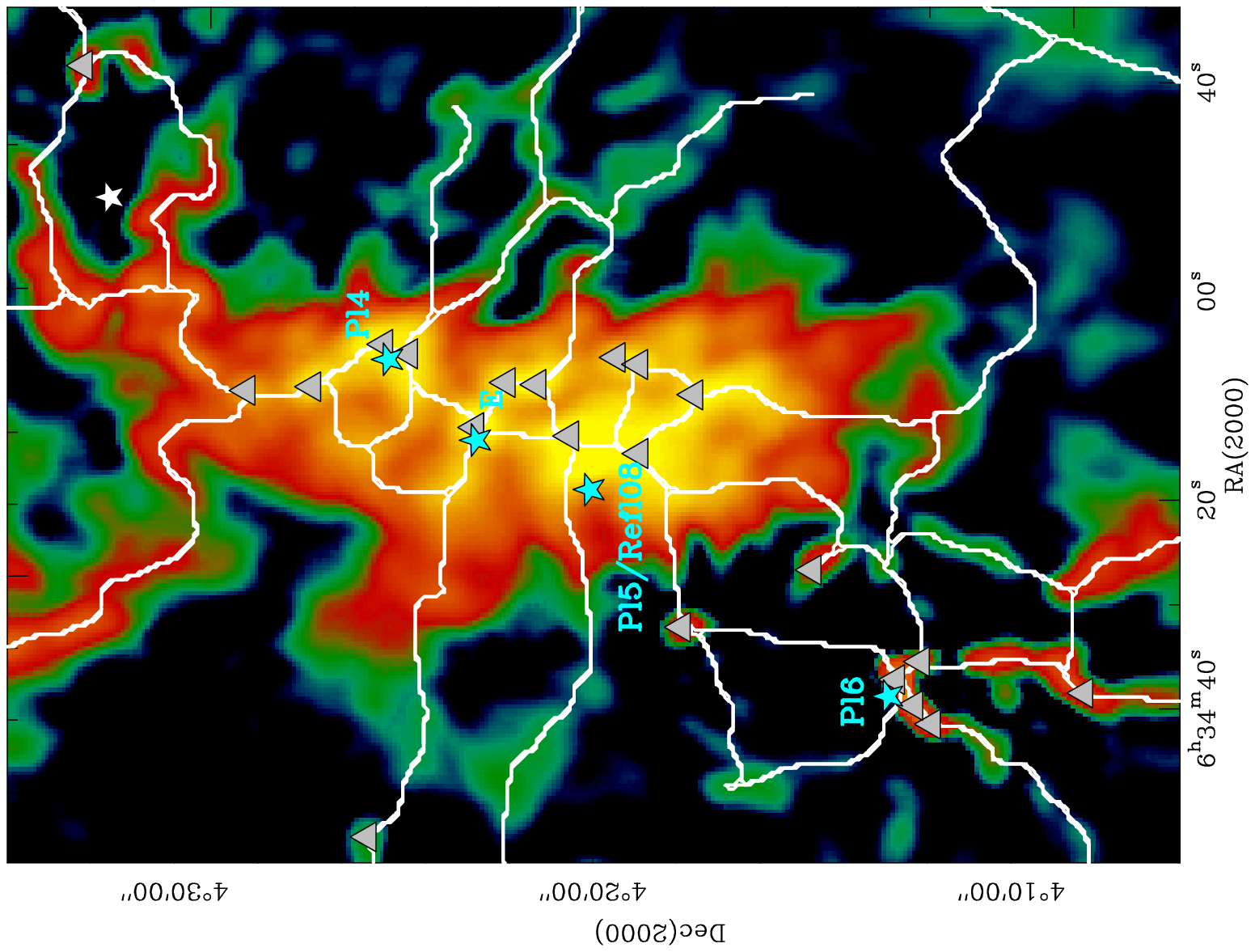}
\end{center} 
\caption [] 
{Close-up of the curvelet image (Fig. 1) of the center region of the
Rosette molecular cloud with the filamentary structure traced by
DisPerSE, indicated with white lines. Different thresholds (called
'persistence', i.e. intensity contrast level, for filament detection
were used (2.9, 1.0, 0.5$\times$10$^{21}$ cm$^{-2}$ from left to
right). The curvelet image has a maximum value for the column density
of 2$\times$10$^{22}$ cm$^{-2}$ with a sigma of 0.5$\times$10$^{21}$
and the original column density map a maximum of 9.4$\times$10$^{22}$
cm$^{-2}$ with a sigma of 1.4$\times$10$^{21}$. }
\label{tests}
\end{figure*}
} 

\section{Curvelet/wavelet decomposition and filament tracing} \label{dis}

There are two parameters determining the detection of filaments:
First, the seperation into curvelets and wavelets, and second the
threshold of DisPerSE to detect filaments. From our experience on the
curvelet/wavelet analysis from Herschel column density maps so far
(Andr\'e et al. 2010; Arzoumanian et al. 2011; Hill et al. 2011), and
from several tests with Rosette, we arrived at a good compromise of
20\% of the intensity being in the curvelets (a difference of around
$\pm$10\%, however, does not change the overall picture). This reveals
the filamentary structure without completely suppressing the more
compact sources. The DisPerSE algorithm detects filaments starting
from a given threshold (defined as difference between saddle points
and peak values) on the curvelet image. However, the column density
map is a 2D-projection of the volume density while DisPerSE works
topologically, connecting all emission features such that projection
effects may create links between filaments that are not physically
related. To overcome this caveat, filament tracing using molecular
line data cubes can be a solution, as first tests on the DR21 filament
have shown (Schneider et al., in prep.).

Figure~\ref{tests} shows as an example a close-up of the crowded
center region of Rosette where different thresholds were applied. A
threshold is defined by an intensity contrast between pixels, the
lowest one starting at 0.5$\times$10$^{21}$ cm$^{-2}$ and accordingly finding many
filaments, up to to a fairly conservative value of
2.9$\times$10$^{21}$ cm$^{-2}$, leaving only the most prominent
features. For this paper, a threshold of 1.0$\times$10$^{21}$
cm$^{-2}$ was choosen in order not to detect too faint filaments. It
is only the most prominent ones that are able to provide enough
material that is accreted onto the cluster center to build up high enough
masses. Again, changing the threshold does not alter these prominent
filaments, they always remain detected.

\section{Probability density function of the Rosette cloud} \label{dis}

The probability density function of the whole cloud obtained from the
column density map (Fig. 2 or 4) is shown in Fig. 6 in linear and
logarithmic scaling. It displays a log-normal form for lower
column densities and a clearly defined power-law tail for higher
column densities that was fitted with a power-law. This is not
generally the case, while the PDF of the high-mass SF region NGC6334
also shows a clear power-law tail (Russeil et al., in prep.), the PDFs
of other high-mass SF clouds are more complex and show several breaks
in the PDFs (Hill et al. 2011, Csengeri et al., in prep.). The reason
for that can partly be line-of-sight effects and limited angular
resolution. A more detailed disussion of PDFs of intermediate- and
high-mass SF regions and comparison to models is presented in Csengeri
et al. (in prep.).

\onlfig{6}{
\begin{figure}[ht]
\begin{center} 
\includegraphics [width=8cm]{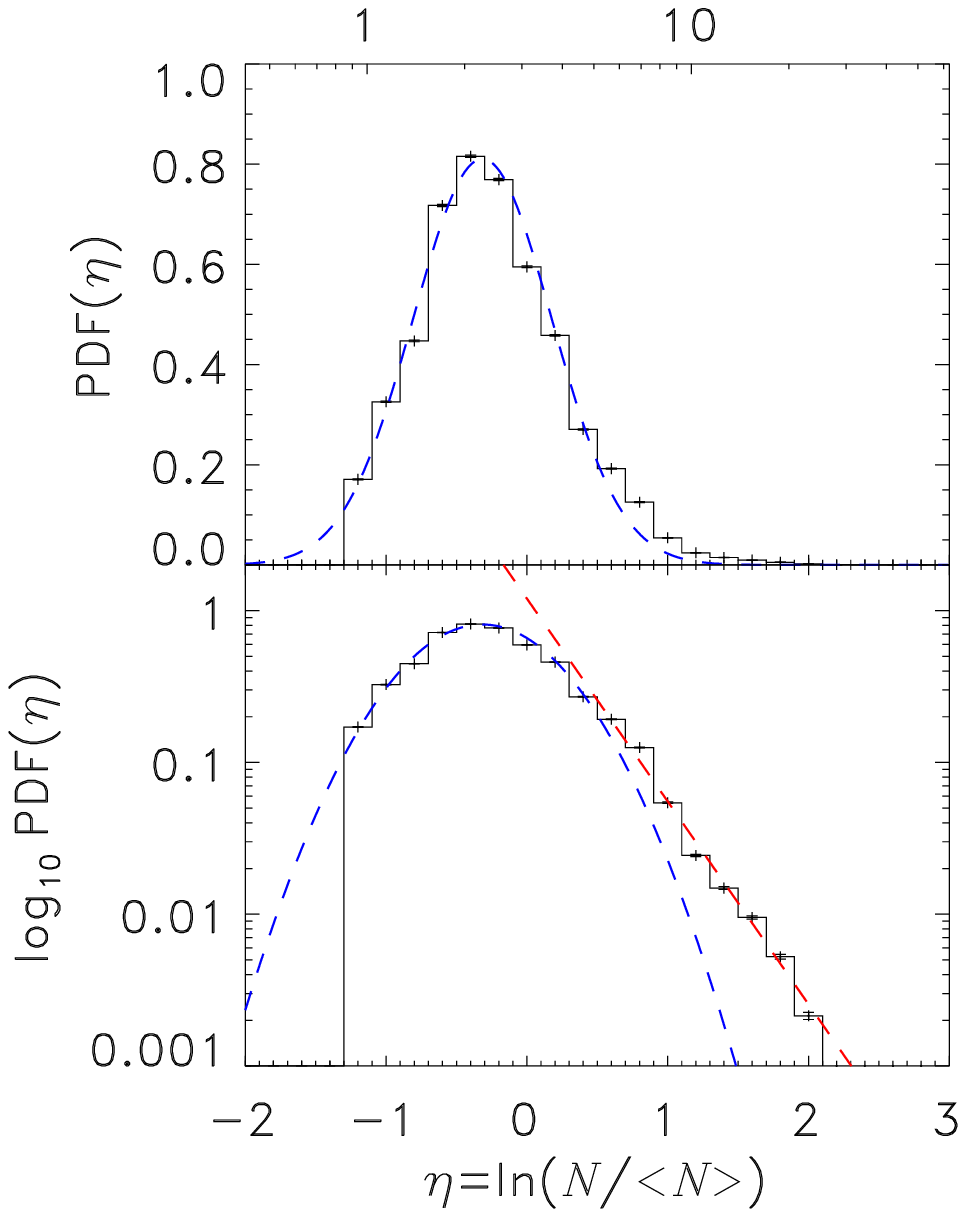}
\end{center} 
\caption [] 
{Probability density function of the whole cloud obtained
from the column density map in linear (top) and logarithmic (bottom)
scaling.  The upper labeling indicates the visual extinction. The red
dashed line shows a power-law fit (the high-density range beyond \av\,=20$^m$  
is not well resolved, we therefore do not attempt to fit a second power law.) }
\label{rosette-total-pdf}
\end{figure}
}

\end{appendix} 

\end{document}